# High-precision measurement of a propagation loss of single-mode optical waveguides on lithium niobate on insulator

**Jintian Lin,**[1,6] **Junxia Zhou,**[2,3,6] **Rongbo Wu,**[1,4] **Min Wang,**[2,3] **Zhiwei Fang,**[2,3] **Wei Chu,**[1] **Jianhao Zhang,**[1,4] **Lingling Qiao,**[1] **and Ya Cheng**[1,2,3,4,5,*]

[1]State Key Laboratory of High Field Laser Physics, Shanghai Institute of Optics and Fine Mechanics, Chinese Academy of Sciences, Shanghai 201800, China.
[2]State Key Laboratory of Precision Spectroscopy, East China Normal University, Shanghai 200062, China.
[3]XXL—The Extreme Optoelectromechanics Laboratory, School of Physics and Materials Science, East China Normal University, Shanghai, China.
[4]University of Chinese Academy of Sciences, Beijing 100049, China.
[5]Collaborative Innovation Center of Extreme Optics, Shanxi University, Taiyuan, China.
[6]Jintian Lin and Junxia Zhou contributed equally to this work.
*Corresponding author: ya.cheng@siom.ac.cn

**We demonstrate fabrication of single-mode optical waveguides on lithium niobate on insulator (LNOI) by optical patterning combined with chemo-mechanical polishing. The fabricated LNOI waveguides have a nearly symmetric mode profile of a mode field size of ~2.5 μm (full-width at half maximum). We develop a high-precision measurement approach by which the single mode waveguides are characterized to have a propagation loss of ~0.042 dB/cm.**

The pursuit of photonic integrated circuit (PIC) has been lasting in the past a few decades as inspired by the enormous success of the electronic integration [1,2]. One of the key building blocks for realizing PICs is the single mode optical waveguides by which transportation and manipulation of photons can be realized in compact and stable optical networks. The major requirements for the waveguides are the low optical loss and high tunability as well as nonlinearity. To this end, crystalline lithium niobate becomes almost an ideal candidate for its broad transparent window and high eletro-optic and nonlinear coefficients [3,4].

Early investigations in establishing PICs on lithium niobate (LN) focused on Titanium diffused LN waveguides with only subtle refractive index contrast, of which the large radius of bending poses a challenge for scalability [3]. On the other hand, several attempts were made on fabricating high quality nanophotonic structures on LNOI with either optical lithography or diamond dicing. The results were not promising owing to the high surface roughness on the sidewalls of the fabricated structures [5,6]. The problem was tackled by employing focused ion beam (FIB) writing to realize a surface roughness of a few nanometers on the sidewalls of nanophotonic structures, resulting in LN microresonators with quality (Q) factors on the order of $10^6$ [7-10]. Afterwards, combinations of either ultraviolet lithography or electron beam lithography with reactive ion etching produced similar results on LNOI [11-13]. Nevertheless, the nanometer-scale roughness left behind by the dry ion etching is still an obstacle to further reduce the optical loss of the LNOI waveguide to below $10^{-1}$ dB/cm. In comparison with the intrinsic absorption loss of the LN crystal which is on the order of $10^{-3}$ dB/cm, the current optical loss can be improved by two orders of magnitude if the scattering loss at the surface can be substantially eliminated. This provides a strong incentive to further suppress the surface roughness on the sidewalls [14].

Recently, we have developed a fabrication approach which has allowed to fabricate both high quality (Q) factor LN microdisks and low loss LN ridge waveguides [15-17]. The technique begins with femtosecond laser micromachining for patterning a hard chromium (Cr) mask coated on the LNOI, followed by chemo-mechanically polishing (CMP) the LNOI sample to transfer the generated pattern to the LNOI. The waveguide fabricated in this manner shows a surface roughness of $R_q$~0.452 nm and a propagation loss of 0.027 dB/cm [16,17]. Unfortunately, the LN waveguides do not support single-mode propagation owing to its relatively large cross-sectional dimensions, which is typically on sub-micrometer but not ~100-nm scale. Here, we convert the multi-mode LN waveguides to single-mode waveguides by covering them with a cladding layer, and examine the mode profiles as well as the propagation losses in the single-mode waveguides.

A commercially available X-cut LNOI wafer was used in fabrication of the ridge waveguides. The LNOI wafer was produced by crystal ion slicing, and then bonded onto a silica layer of a 2 μm-thickness supported by a bulk LN substrate of a 500 μm-thickness. The structure of LNOI wafer is shown in Fig. 1(a). The ridge waveguide was oriented along Y-axis. In such a configuration, the transverse electric (TE) mode in the waveguide experiences the extraordinary refractive index of the LN crystal. The process flow for fabricating the ridge waveguides includes five steps as illustrated in Fig. 1.

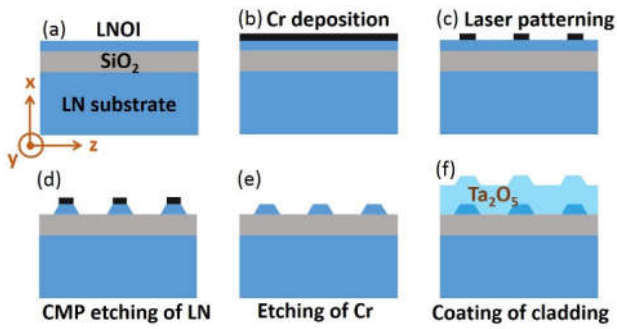

Fig. 1. The process flow of LNOI waveguide fabrication.

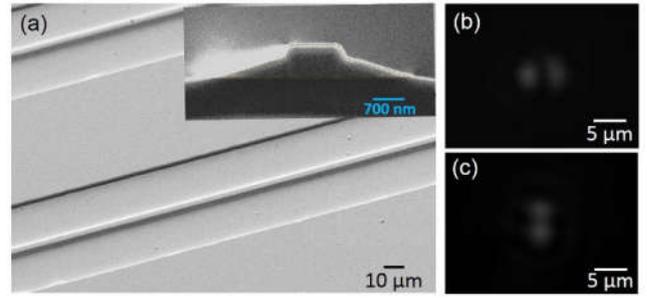

Fig. 2 (a) SEM images of the LN waveguide before coating the Ta2O5 cladding layer). Inset in (a): cross section of the waveguide. The spatial distributions of (b) TE and (c) TM (transverse magnetic) modes.

First, a wear-resisting Cr layer with a thickness of 600 nm was coated on LNOI wafer by magnetron sputtering, as shown in Fig. 1(b). Note that the hardness of Cr is much higher than that of LN, making Cr a good hard mask material for protecting the underneath LN in the CMP process. Second, the Cr layer was patterned into stripes with a width of ~1 μm by femtosecond laser ablation, as shown in Fig. 1(c). The femtosecond laser ablation was performed at a repetition rate of 250 kHz and an average power of 0.05 mW. An objective lens (Model: M Plan Apo NIR, Mitutoyo Corporation, Japan) with a numerical aperture (NA) of 0.7 was used to focus the laser pulses, creating a focal spot of a diameter of ~1 μm on the sample. This objective lens was mounted on a one-dimensional (1D) motion stage (Model: ANT130-110-L-ZS, Aerotech Inc., USA), which travels in the vertical (Z) direction at a resolution of 100 nm to ensure accurate focusing onto the sample surface. The sample was mounted on an XY stage (Model: ABL15020WB and ABL15020, Aerotech Inc., USA) with a translation resolution of 100 nm. The Cr was patterned by scanning the focal spot across the areas according the designed patterns. All the motion stages were computer programmable. The laser power we chose was sufficient for ablating Cr, but insufficient for ablating LN crystal, because the damage threshold of LN is higher than that of Cr [18]. This characteristic ensures that in the ablation of Cr, the LN thin film keeps intact. Afterwards, the LNOI wafer was subjected to the CMP which was carried out using a wafer polishing machine. A LN ridge structure was obtained after the CMP as shown in Fig. 1(d). A smooth sidewall with an average roughness of ~0.5 nm was attainable by CMP. The Cr mask was removed by chemical etching by immersing the sample into a Cr etching solution for 4 min, as shown in Fig. 1(e). Lastly, a layer of Ta2O5 was coated on the sample to create a suitable refractive index contrast for ensuring single-mode waveguiding, as shown in Fig. 1(f).

The scanning electrical microscope (SEM) image of the fabricated LN ridge waveguide is presented in Fig. 2(a), and the cross section of waveguide is shown in the inset. The top width of the waveguide was determined to be ~1.0 μm, whereas the bottom width was measured as ~4.2 μm. The spatial modes were excited and characterized by coupling the waveguide with 1550 nm-wavelength laser using a fiber lens. The polarization state of the input light was controlled by an in-line polarization controller. The light transmitted from the output port of the waveguide was collected by an objective lens with NA=0.3. A beam expander was introduced behind the objective lens, and projects the image of the output port on an infrared CCD (InGaAs camera, HAMAMATSU Inc.). The spatial distributions of TE and TM modes were captured by the CCD, as shown in Fig. 2(b) and (c), respectively. The waveguide supports high-order spatial modes for the TE and TM modes owing to the high refractive index of LN and large transverse dimensions of the waveguide.

To produce single mode LNOI waveguides for both TE and TM modes, Ta2O5 (refractive index 2.057) of an ultralow loss [19], was deposited on the fabricated sample by electron beam evaporation. Figure 3(a) shows the SEM image of the LN ridge waveguide covered with the Ta2O5 cladding layer of a thickness of 3.5 μm. A single mode spatial profile was obtained for TE mode as shown in Fig. 3(b), which is consistent with the calculated mode profile in Fig. 3(c). Similarly, such waveguide supports single mode propagation for TM mode as well, as shown in Fig. 3(d) and evidenced by the corresponding calculation result in Fig. 3(e). The full width at half maximum (FWHM) of TE mode is measured to be ~2.5 μm, and the FWHM of TM mode is ~2.3 μm.

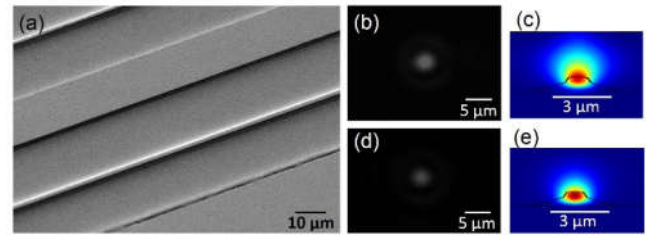

Fig. 3 (a) SEM image of the waveguide covered with Ta2O5. The (b) measured and (c) calculated TE mode profiles. The (d) measured and (e) calculated TM mode profiles.

Traditionally, the propagation loss of ridge waveguide can be measured based on a Fabry-Perot (FP) cavity measurement or direct cut-back methods [6,20-22]. However, the ultralow loss of our chemo-mechanically polished LNOI waveguide cannot be determined by the above methods owing to the insufficient precisions. To overcome this difficulty, we develop a unique technique which allows us to reliably measure the nearly inappreciable loss in the on-chip LNOI waveguide.

Our high-precision loss measurement method is established based on the design illustrated in Fig. 4(a). The device is composed of three beam splitters aligned in a vertical array each of which consists of two identical electro-optic (EO) Mach-Zehnder interferometers (MZIs) bridged by an EO phase shifter. The similar design has been used to produce high extinction ratio beam splitters immune to the imperfections in the fabrication [Refs. 23,24 ]. The beam splitter was fabricated on an X-cut LNOI chip with its optic axis oriented perpendicularly to the MZI arms, as that in Fig. 3(a). The beam splitting ratio of the fabricated directional coupler (see Fig. 4(b)) is designed to be 7:3. The output arms of different lengths were fabricated to differentiate the propagation loss, i.e., one arm is 12 mm longer than the other one. After the fabrication of the LNOI waveguides, Au electrodes with a thickness of 200 nm were added by a magnetron sputtering followed by a space-selective patterning via femtosecond laser ablation, as shown in Fig. 4(c). The gap $d$ between the Au electrodes in each MZI was set as 10 μm, making the two electrodes symmetrically arranged on both sides of the LNOI waveguides. The lengths of the interferometer arms of MZI 1, Phase Shifter, MZI 2, which are sandwiched between the Au electrodes, are 2 mm, 11 mm, and 2 mm, respectively. Photograph of the fabricated device captured by a digital camera is shown in the inset of Fig. 4(b). The total length of the chip is ~ 30 mm.

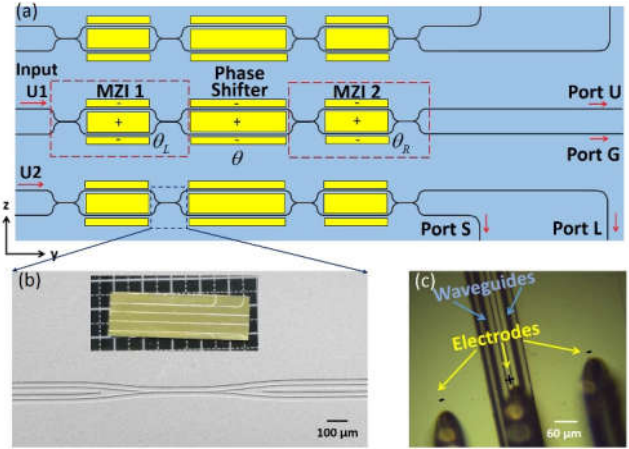

Fig. 4. (a) Layout of the device. Here the phase differences $\theta_L$, $\theta$, and $\theta_R$ are contributed by the fabrication imperfections of the interferometer arms and that of EO modulation. (b) SEM image of the directional coupler, Inset: overview photograph of the fabricated device consisting of three beam splitters. (c) The optical microscope image of the electrodes contacted by 3 pins.

To characterize the EO response of the beam splitter, the telecom laser (New Focus Inc., Model TLB 6728) with a pure TE mode was coupled into the input port U1 of the beam splitter through a fiber lens. The output beam was first collimated by a 50× objective lens (Model: M Plan Apo NIR, Mitutoyo Corporation, Japan) of NA 0.42, and then sent into an optical spectrum analyzer (OSA) (YOKOGAWA Inc., Model AQ6370D, dynamic range 45 dB). With this arrangement, the electric field was parallel to the optic axis of the LN as evidenced in Fig. 5 (a). The phase difference $\phi$ between the waves exiting from the two interferometer arms can be expressed as:

$$\phi = \frac{2\pi}{\lambda} n_e^3 r_{33} \frac{V}{d} l \quad (1)$$

Here, $\lambda$ is the wavelength (i.e., 1550 nm), $n_e$ is the refractive index of extraordinary light, and $r_{33}$ (= 30.8 × 10$^{-12}$ m/V) is the largest electro-optic coefficient of LN, V is the applied voltage. The measured half-wave voltage $V_\pi$ are 6.7 V for the Phase Shifter and 36.8 V for MZI 1 and MZI 2. The results agree well with the numerical calculation. Next, we tuned the splitting ratio for both MZI 1 and MZI 2 to a precise 50:50 using the following procedures [Ref. 23]:

1. Adjust the voltage on the Phase Shifter to minimize the output power of output port G.
2. Scan the voltages on both MZI 1 and MZI 2 to minimize the output power of port G.
3. Repeat Steps (1)-(2) if necessary until the minimum power in the output G is zero, and the maximum power in the other port (i.e, U) is as large as possible.

In our experiment, the extinction radio between port G and port U was determined to be ~40 dB based on the experimental curve in Fig. 5 (b). The curve in Fig. 5 (b) was obtained by varying the phase difference $\theta$, giving rise to an oscillating power curve expressed by [23]:

$$P_U = \frac{1}{2}(1 - \cos\theta) \quad (2)$$

The measured output power at port U in Fig. 5 (b) nicely follows the cosine curve obtained using Eq. 2, and that at port G remains sinusoidal.

To measure the propagation loss of the single mode waveguide, the laser was coupled to the input port U2 of the bottom beam splitter in Fig. 4(a), and both MZI were tuned at 50:50 splitting ratio via EO modulation. Then the power output from Port S was tuned to be maximum via adjusting the voltage applied on phase shifter. Finally, the power output from Port S was tuned to be the half of the maximum value. In other words, the phase difference $\theta$ between the two arms of Phase Shifter in the bottom beam splitter was $\pi/2$, and the powers injected into the output ports S and L are the same. The propagation loss was measured to be 0.042 dB/cm by comparing the powers at the output port S (counts: 3943) and L (counts: 3898).

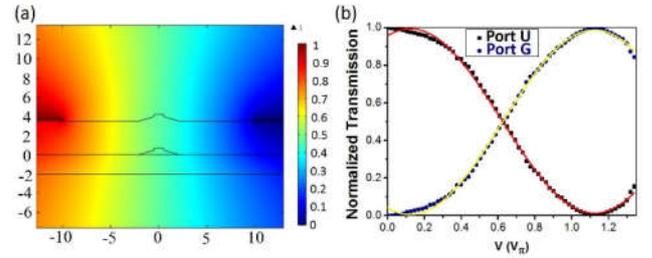

Fig. 5 (a) The calculated normalized electric field distribution in the cross sectional plane of the LNOI waveguide, showing that the electric field is almost parallel to the optic axis of LN. (b) Normalized transmission spectra of the output ports U and G as a function of the applied voltage V, featuring a sinusoidal-like curve.

To conclude, we fabricate single mode LNOI waveguides with a propagation loss of ~0.042 dB/cm and a mode field size of ~2.5 μm. The high-precision loss measurement is achieved using an EO controllable beam splitter to ensure the simultaneous injection of two light waves of a same input power into two waveguides of unbalanced lengths. This method avoids the fluctuation in the coupling efficiency when carrying a conventional cut-back measurement, enabling to differentiate two waveguides of propagation losses close to each other. The low-loss single mode LN waveguides can be used for constructing complex photonic circuits.